# Near Linear OS Scheduling Optimization for Memory Intensive Workloads on Multi-socket Multi-core servers

**Murthy Durbhakula**

**Indian Institute of Technology Hyderabad, India**

**cs15resch11013@iith.ac.in, murthy.durbhakula@gmail.com**

**Abstract.** Multi-socket multi-core servers are used for solving some of the important problems in computing. Remote DRAM accesses can impact performance of certain applications running on such servers. This paper presents a new near linear operating system (OS) scheduling algorithm to reduce the impact of such remote DRAM accesses. By keeping track of the number of local and remote DRAM accesses, using performance counters, for every thread and applying this algorithm, I come up with a new schedule of threads for the next quantum. This new schedule reduces remote DRAM accesses and improves overall performance. I also show that this algorithm is actually linear in the best case. As the algorithm is near-linear it is amenable for implementation in a real operating system.

## 1 INTRODUCTION

Cache coherent NUMA (ccNUMA) multi-socket multi-core servers today are used for solving some of the important problems in computing. Performance of memory-intensive workloads on such servers are impacted by remote DRAM accesses. One way to solve this problem is to rewrite the application. An alternative way is to observe remote DRAM access patterns at run-time using performance counters and optimize the OS scheduler to schedule threads in a way to minimize the impact of these accesses. In this paper I present such an OS scheduling optimization, a near linear algorithm, which keeps track of local and remote DRAM accesses for every thread and then applies the optimization to decide to which socket to schedule each thread for the next scheduling quantum. To the best of my knowledge, none of the current commercial operating systems optimize their scheduler for remote DRAM accesses. In Section 2 I present the near linear scheduling algorithm and in Section 3 I conclude the paper.

## 2 SCHEDULING ALGORITHM

For the algorithm I present in this section I assume that there are performance counters available that can keep track of number of DRAM accesses made by every thread to every socket.

**Near linear Algorithm**

For an operating system scheduling algorithm, on a context switch, we have to decide the next schedule in as little time as possible. Near linear algorithm shown below has better algorithmic complexity and hence is more amenable to be implemented in a real operating system. As part of future work I plan to quantitatively compare this algorithm other algorithms using real benchmarks. In a four socket system, first, Socket0 will find top four threads with highest number of DRAM accesses. In parallel, Socket1 will find top eight threads with highest number of DRAM accesses, and Socket2 will find top 12 threads with highest number of DRAM accesses. All this can be done in linear time using Order Statistics [1].Top four threads found by Socket0 are assigned to Socket0. Then Socket0 will communicate four thread numbers to Socket1. If four thread numbers are all present in top eight thread numbers in Socket1 then we can assign remaining four of eight threads to Socket 1. Otherwise Socket1 will remove threads assigned to

Socket 0 and calculate top four threads out of remaining and assign those to itself. Then Socket1 will communicate all eight thread numbers, which are already assigned to Socket 0 and Socket1, to Socket2. If all eight thread numbers are in top 12 threads, Socket2 will assign remaining four in top twelve to itself. Otherwise it will remove those eight threads which are already assigned and calculate top four from the remaining and assign to itself. Now socket2 will communicate all 12 threads, assigned to Socket0, Socket1, and Socket2, to Socket 3. Socket3 will assign remaining four of 16 threads to itself. The worst case complexity is calculating top four threads as many times as there are sockets. Hence it is O(NK) where N is number of threads and K is number of sockets. In the best case where none of the sockets has to calculate top four threads again the algorithmic complexity is actually linear in number of threads O(N).

---

Input: Threads T0,…TN with DRAM accesses to sockets N0…NL. Each socket has 4 cores and can run one thread per core. Present schedule S_present which has a mapping of N threads to L sockets.

Output: New schedule S_next with new mapping of threads to sockets.

begin

1. Every socket has N DRAM access counts from N threads. Say there are four sockets and four threads per socket. Socket0 will find top four DRAM access counts. Socket1 will find top eight DRAM access counts. Socket2 will find top 12 DRAM access counts. Socket3 will also find top 12 DRAM access counts. We dont need sorted output. Hence this can be done in linear time using Order Statistics.

2. Top four threads found by Socket0 are assigned to Socket0. Then Socket0 will communicate four thread numbers to Socket1. If four thread numbers are all present in top eight thread numbers in Socket1 then we can assign remaining four of eight threads to Socket 1. Otherwise Socket1 will remove threads which are already assigned and calculate top four threads among remaining and assign those to itself. Then Socket1 will communicate all eight thread numbers, which are already assigned to Socket 0 and Socket1, to Socket2. If all eight thread numbers are in top 12 threads Socket2 will assign remaining four in top twelve to itself. Otherwise it will remove threads which are already assigned and calculate top four out of remaining and assign to itself. Now socket2 will communicate all 12 threads, assigned to Socket0, Socket1, and Socket2, to socket 3.
Socket3 will assign remaining four of 16 threads to itself. The worst case complexity is calculating top four threads as many times as there are sockets. Hence it is O(NK) where N is number of threads and K is number of sockets. Generally number of sockets tend to be small and we see a growing trend in number of threads-per-socket. Hence this algorithm is near-linear.

end

Overall complexity of the algorithm is O(NK).

**Near Linear Algorithm**

## 3 CONCLUSION

Cache coherent NUMA (ccNUMA) multi-socket multi-core servers today are used for solving some of the important problems in computing. Performance of memory-intensive workloads on such servers are impacted by remote DRAM accesses. I presented a new near linear OS scheduling algorithm that observes DRAM accesses made by every thread to every socket and uses that information to optimally schedule threads for the next scheduling quantum thus reducing remote DRAM accesses. As part of future work I plan to quantitatively compare this algorithm with other algorithms using real benchmarks.

## Acknowledgements

I would like to thank Prof. Alan Cox of Rice University for initially discussing with me the concept of optimizing OS scheduling algorithms for improving the performance of various workloads. I would also like to thank various reviewers of earlier version of this work for their comments and feedback. Finally I would like to thank my wife and kids for supporting me morally during the course of this work.